\documentclass[sigconf]{acmart}

\usepackage{subcaption}
\usepackage{cleveref}

\usepackage{algorithm}
\usepackage{algpseudocode}

\algrenewcommand\algorithmicindent{0.8em}%

\usepackage[frozencache]{minted}

\usepackage{listings}

\usepackage{lipsum}

\usepackage{tabularx}

\usepackage{booktabs}
\usepackage{multirow}

\usepackage{enumitem}

\usepackage{multicol}

\usepackage[subtle]{savetrees}

\newcommand{\theinstitution}{Federico II University}

\AtBeginDocument{%
  }

\copyrightyear{2025}
\acmYear{2025}
\setcopyright{rightsretained}
\acmConference[SIGCSE TS 2025]{Proceedings of the 56th ACM Technical Symposium on Computer Science Education V. 1}{February 26-March 1, 2025}{Pittsburgh, PA, USA}
\acmBooktitle{Proceedings of the 56th ACM Technical Symposium on Computer Science Education V. 1 (SIGCSE TS 2025), February 26-March 1, 2025, Pittsburgh, PA, USA}
\acmDOI{10.1145/3641554.3701836}
\acmISBN{979-8-4007-0531-1/25/02}

\makeatletter
\gdef\@copyrightpermission{
   \begin{minipage}{0.3\columnwidth}
     \href{https://creativecommons.org/licenses/by-nc-sa/4.0/}{\includegraphics[width=0.90\textwidth]{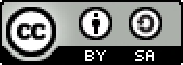}}
   \end{minipage}\hfill
   \begin{minipage}{0.7\columnwidth}
     \href{https://creativecommons.org/licenses/by-nc-sa/4.0/}{This work is licensed under a Creative Commons Attribution-ShareAlike International 4.0 License.}
   \end{minipage}
   \vspace{5pt}
}
\makeatother

\begin{document}

\title[Evaluation of Systems Programming Exercises through Tailored Static Analysis]{Evaluation of Systems Programming Exercises\\ through Tailored Static Analysis}

\author{Roberto Natella}
\orcid{0000-0003-1084-4824}
\affiliation{%
  \institution{Università degli Studi di Napoli Federico II}
  \city{Napoli}
  \state{}
  \country{Italy}
}
\email{roberto.natella@unina.it}


\begin{abstract}
In large programming classes, it takes a significant effort from teachers to evaluate exercises and provide detailed feedback. In systems programming, test cases are not sufficient to assess exercises, since concurrency and resource management bugs are difficult to reproduce. This paper presents an experience report on static analysis for the automatic evaluation of systems programming exercises. We design systems programming assignments with static analysis rules that are tailored for each assignment, to provide detailed and accurate feedback. Our evaluation shows that static analysis can identify a significant number of erroneous submissions missed by test cases.
\end{abstract}





\maketitle

\section{Introduction}

Programming exercises are a key aspect of computer science education \cite{huet2004new,ramirez2015increasing,lee2021feedback}, as they engage students thorough the semester and in preparation for the exams \cite{braught2021pair,ahadi2016performance}. 
To make the most from practical exercises, it is important that students receive detailed feedback on their programs. However, teachers must make a significant effort to design the exercises, collect submissions, evaluate and annotate them with detailed feedback, and distribute feedback. Moreover, it is often necessary to perform multiple rounds of evaluation, as students may incur new errors or incorrectly fix the previous ones. The problem is further exacerbated by the large number of students in programming classes, and by the need for multiple exercises during a semester.

Therefore, it is important to \emph{automate} the evaluation process in order to reduce the effort to a reasonable level, and to enable teachers to run more practical exercises for large classes. The most common approach is to provide students with template code (e.g., a skeleton of the program) and \emph{test cases} written by teachers, in order to evaluate the correctness of the final program. An example of this approach is implemented in the popular GitHub Classroom \cite{hecht2023github}.

However, this approach is not effective for \emph{systems programming} exercises \cite{kay2003unix,kerrisk2010linux}, i.e., programs that are concurrent (e.g., multi-threaded) and that make use of low-level resources of the OS, such as semaphores, message queues, shared memory, and others. Moreover, these exercises are written in C language, where the programmer must handle raw memory and manage its allocation. These features make it difficult to automatically evaluate correctness by means of test cases. In particular, concurrency bugs (e.g., race conditions and deadlocks, due to missing or wrong use of synchronization primitives) are notoriously difficult to detect by running a program, since they only cause failures under rare combinations of events and CPU scheduling \cite{grottke2016mandelbugs}. Similarly, resource management bugs (e.g., leaks of OS resources) lead to side effects that are not easily found by looking at the behavior of a program. Finally, test cases do not provide detailed feedback about which part of the source code is buggy.

This paper presents an experience report on the use of \emph{static analysis} for the automatic evaluation of systems programming exercises. 
Static analysis can identify concurrency and resource management bugs in a reliable and accurate way, since it inspects the contents of the program with actually running it, and it is independent of the timing of events and CPU scheduling. Thus, static analysis represents an appealing solution to the limitations of dynamic analysis (i.e., test cases). The goal of this paper is to show that the use of static analysis is feasible and useful. In particular, static analysis has been \emph{tailored} for each assignment, in order to provide feedback that is detailed (i.e., indicating the faulty code, with an explanation) and accurate (i.e., not prone to false alarms). The paper presents this approach in the context of a class on Operating Systems at the \theinstitution, by discussing the design of systems programming exercises and how to define tailored static analysis rules. This approach was experimented on a large class, using an automated pipeline to collect and analyze submissions and distribute feedback. All code has been released as open-source software. Our experience shows that the approach is able to identify a significant amount of bugs that are not otherwise detected by test cases.

In the following, Sec.~\ref{sec:related_work} discusses related work. Sec.~\ref{sec:methodology} presents our approach. Sec.~\ref{sec:case_study} presents the case study. Sec~\ref{sec:conclusion} concludes the paper.
\section{Related Work}
\label{sec:related_work}

Automatic evaluation is typically performed by running the submitted code with inputs from test cases, and checking that the outputs comply with the outputs expected by the test cases. For example, systems that adopt this approach are ArTEMiS \cite{krusche2018artemis} and VPL \cite{zampirolli2021experience,rodriguez2012virtual}; SAUCE \cite{hundt2017sauce} applied this paradigm in parallel programming. Galan et al. \cite{galan2019automated} report on the systematic application of this approach in multiple courses over several years. Paiva et al. \cite{paiva2022automated} presented a systematic review of studies on automatic assessment in computer science education, including an analysis of techniques for testing and generating feedback.

Paiva et al. recognize that static analysis techniques are a promising direction for automatic assessment, although the studies in this area are still limited. Among them, ASSYST \cite{jackson1997grading} statically analyses programs by computing complexity metrics (e.g., McCabe cyclomatic complexity), style metrics (module length, number of comment lines, use of indentation), and performance metrics using a code profiler for counting statement executions. However, these metrics are only an indirect proxy for code quality. Algo+ \cite{bey2018comparison} assesses programs by comparing them to a set of predefined solutions selected from past correct and erroneous submissions, which were already assessed by an instructor. A limitation of this approach is that comparisons can still be inaccurate, and do not provide indications on which specific part of a program is faulty. Aziz et al. \cite{aziz2015parallel} presented a plugin for Web-CAT \cite{edwards2008web} to assess parallel programming assignments, by integrating off-the-shelf static analysis tools (Checkstyle and FindBugs) to identify common style and programming mistakes for the Java language. Similarly, SOBO \cite{bobadilla2023sobo} uses SonarQube as a basis to automatically generate feedback. Even if static analysis tools are quite powerful, their use has been limited to their default configuration, by running general-purpose queries that are unaware of the requirements and expectations for the program. Therefore, they are prone to false positives and false negatives, which is a general problem of static analyzers \cite{chess2007secure}. Moreover, Senger et al. \cite{senger2022static} found that warnings from static analysis tools can be inversely correlated with correctness.


This paper adopts a unique approach to static analysis. Our approach leverages recent advances in static analysis, which allows the user to write new, tailored queries using domain-specific languages. Tools of this kind include Semgrep \cite{semgrep}, CodeQL \cite{codeql}, and Joern \cite{joern}. We designed a set of systems programming exercises and related checks to identify correct uses of synchronization patterns and of resource management APIs. Since our static analysis rules are tailored for the assignment, they are aware of which are the correct APIs and their parameters that should be used for each specific context.

\section{Methodology}
\label{sec:methodology}

\begin{figure}[!ht]
\includegraphics[width=\columnwidth]{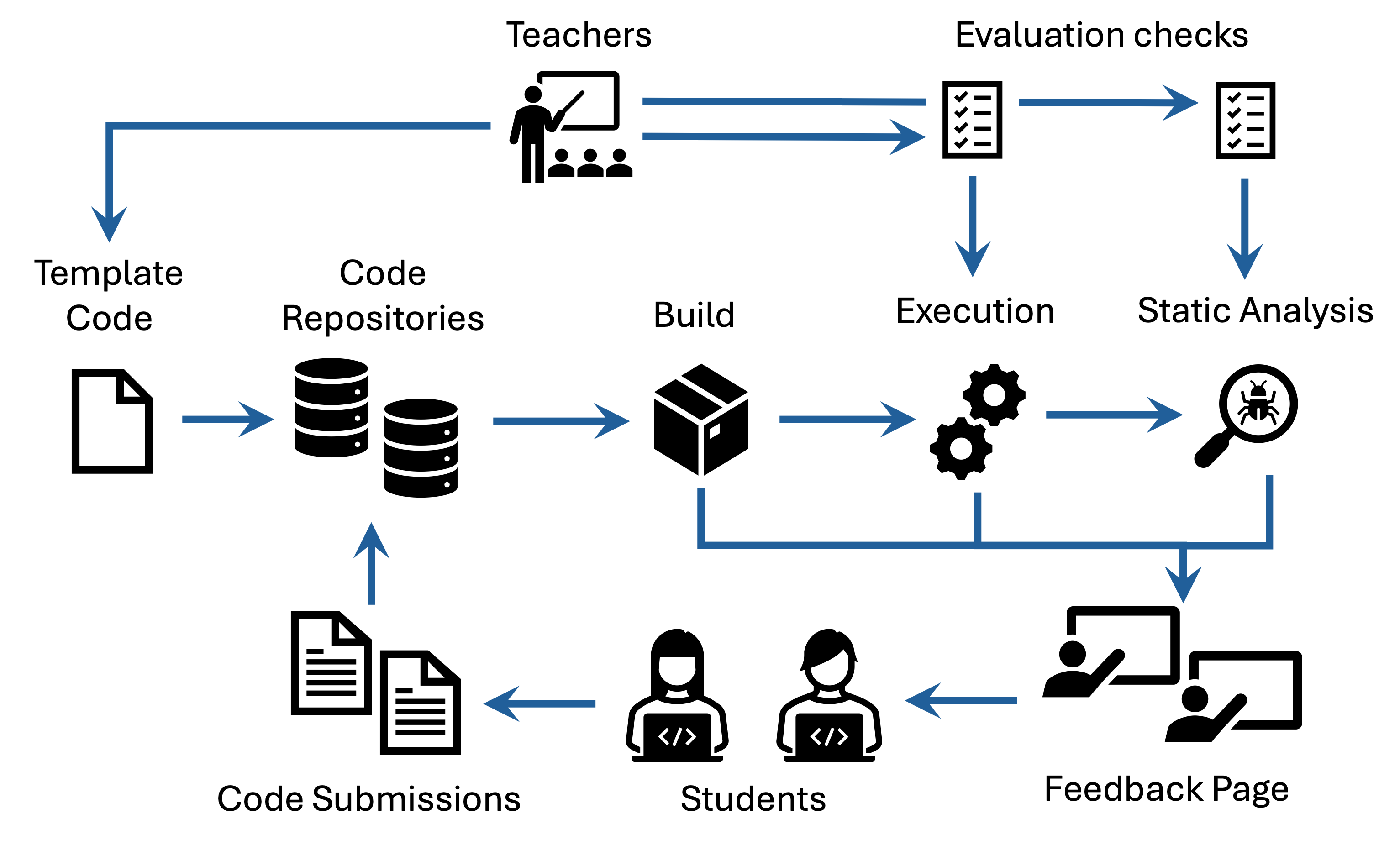}
\vspace{-0.7cm}
\caption{Overview of the code evaluation process.}
\label{fig:overview}
\end{figure}

The proposed approach is shown in Figure~\ref{fig:overview}. First, the teacher prepares a programming assignment, which includes (1) a brief description of the objective of the program, and (2) a template code, which the students will  extend for the assignment. The template contains the basic structure of the program, including files, function prototypes, and  business logic without any synchronization or resource management. Moreover, the template contains ``\emph{TBD}'' (\emph{to be developed}) comments, which provide detailed indications about the expectations of the assignment. 
In addition to the assignment, the teacher also prepares \emph{evaluation checks} for automatic assessment. The checks are in the form of small scripts and configuration files.

The assignment is then distributed among the students, by creating \emph{code repositories} with the template code and a description of the assignment in a README file. Repositories can be managed with automated systems, such as GitLab \cite{gitlab} or Gitea \cite{gitea}, using Git for version control \cite{lee2021feedback}. We released a set of open-source scripts to automate the creation of repositories and accounts for each student, on a GitLab or Gitea server \cite{git-esami}. GitHub Classroom \cite{hecht2023github} is another option to automate the creation of repositories on \url{github.com}.

Each time a student submits code through a commit, the evaluation process is automatically triggered. The evaluation process consists of three stages: build, execution, and static analysis. If the submitted code does not pass any of these three stages, the process is interrupted and a feedback message for the student is posted on a web page, such as, on the issue tracker for the repository on \url{github.com}. The feedback message provides information on the evaluation check that found the issue. Moreover, for static analysis, the feedback message includes the location of the faulty code. The student can then fix the program and make a new submission. This process provides timely feedback for each submission through automation, and can be iterated indefinitely.

The build stage is performed using a build automation tool, such as Make, which compiles the submitted code using build configuration rules included in the template code. Any build error is reported as feedback to the student. Then, the binary executable of the program is executed, and its output is collected. The execution stage first evaluates whether the program completed the execution, without crashing or stalling. Moreover, the stage evaluates the status of OS resources (such as, UNIX semaphores), both before and after the execution of the program. If the program is correct, it should automatically deallocate OS resources once it terminates. If the execution leaves any OS resource that was not present before the execution, then a resource leak problem is detected and reported to the student. Finally, the execution stage performs checks that are specific for the assignment, by analyzing the output of the program. If the output is not consistent with the expectation, a failure is reported.

Finally, the static analysis stage performs additional checks by analyzing the source of the code submission. The static analysis stage is able to identify issues missed by the execution stage. Many types of bugs in systems programming, such as concurrency and resource management bugs, are unlikely to surface by executing the program. Instead, static analysis does not rely on executing the program, but looks at the source code to check that the student applied the right programming pattern. These checks are tailored for the assignment, by assessing the submission at specific points of the source code. 

A key aspect of this approach is the definition of evaluation checks. The checks that evaluate the build of the program and the occurrence of crashes, stalls, and OS resource leaks are independent of the specific assignment. However, to perform a thorough evaluation of the code submission, we also need assignment-specific checks. 
In the following, we will first present a detailed example of a systems programming assignment (Sec.~\ref{subsec:example}), and its static and dynamic checks for the evaluation. Then, we discuss the general criteria for defining the evaluation checks (Sec.~\ref{subsec:general_criteria}).

\subsection{Example of assignment and evaluation}
\label{subsec:example}

The objective of this assignment is to develop a simulator of an I/O scheduler, which is inspired by the theory lectures of the class. Students have to apply the producer-consumer scheme with a circular buffer, as also taught in the class. The exercise involves a consumer process (the I/O scheduler), and multiple producer processes (applications writing to the disk). The template code includes function prototypes and data structures to represent the circular buffer and the I/O requests to be produced and consumed. Moreover, the template code includes the business logic of the producer and consumer processes, which invokes the functions that actually perform the operations on the ring buffer (\texttt{insert\_request()} and \texttt{pick\_request()}). The business logic provides inputs for the execution of the program.

\begin{figure}[!ht]
  \centering
  \subcaptionbox{Template code\label{fig:prod_cons_tbd}}{%
    \includegraphics[width=\columnwidth]{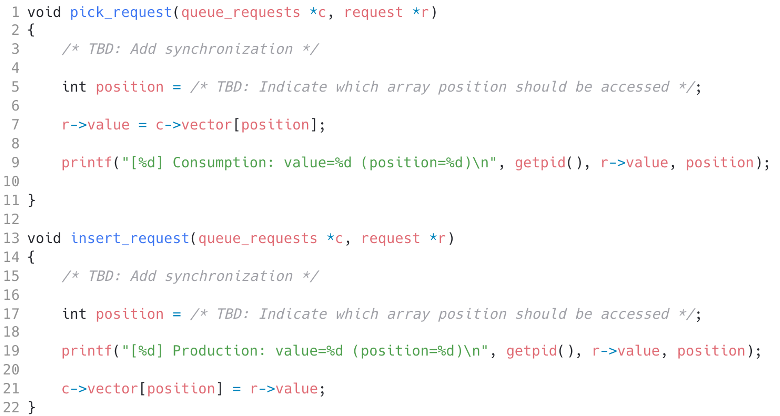}\vspace{-0.3cm}
  }
  \subcaptionbox{Solution\label{fig:prod_cons_solved}}{%
    \includegraphics[width=\columnwidth]{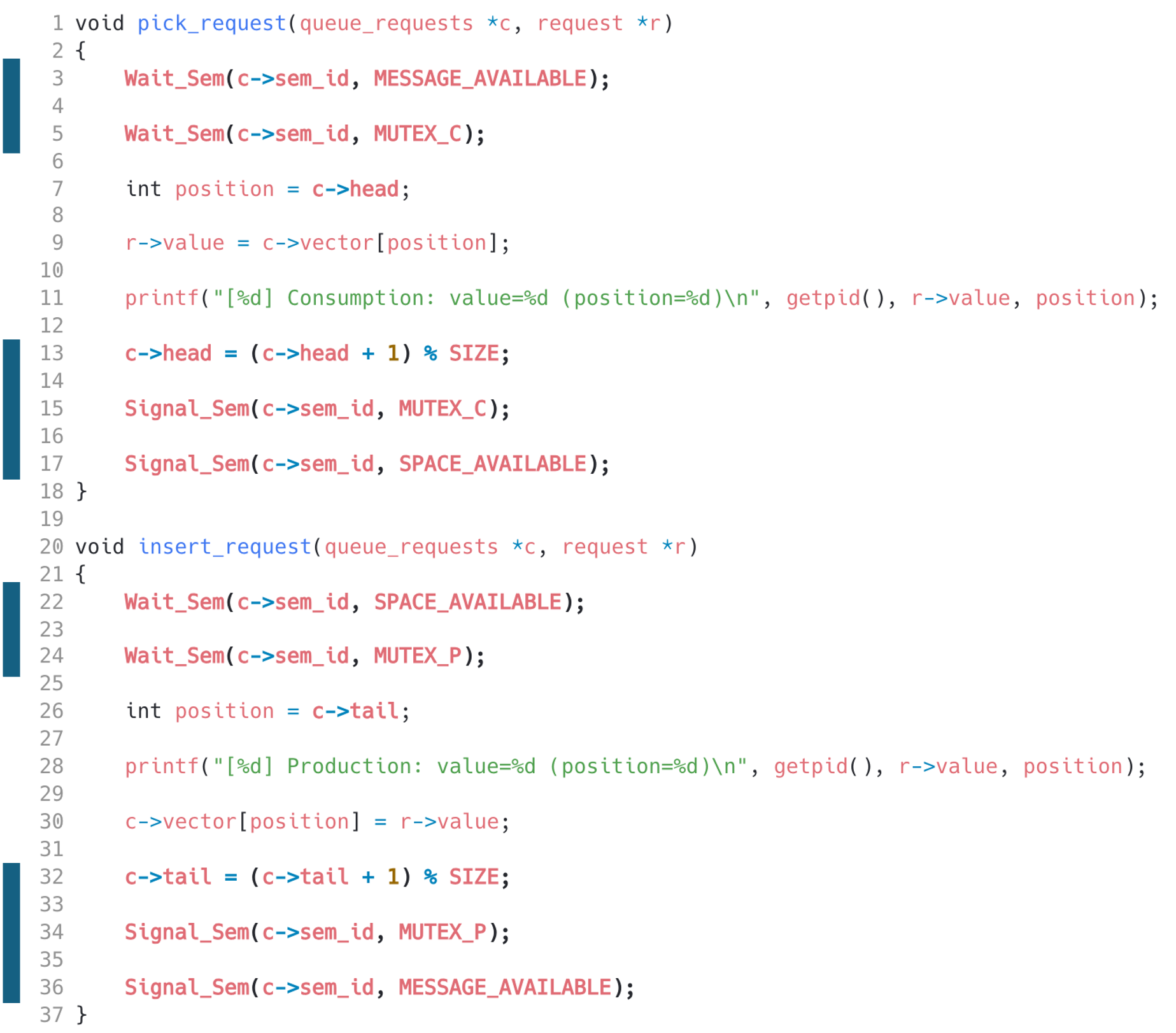}\vspace{-0.3cm}
  }
  \vspace{-0.3cm}
  \caption{Assignment on producer-consumer.}
  \label{fig:example_prod_cons}
\end{figure}

The assignment requires the students to complete the template code, by adding code to create child processes, to manage OS resources (UNIX semaphores and shared memory) needed for the program, and to add code for producer-consumer synchronization. In particular, Figure~\ref{fig:prod_cons_tbd} shows template code for the producer and consumer functions, with ``\emph{to be developed}'' comments to ask students to introduce synchronization in these functions. Figure~\ref{fig:prod_cons_solved} shows the solution for this assignment by using UNIX semaphores. \texttt{Wait\_Sem()}, a convenience function provided with the template code, suspends the producer at line 22 if there is no space available in the ring buffer to insert an I/O request, and at line 24 if there is already another producer that is writing to the ring buffer. A suspended producer can be woken up by \texttt{Signal\_Sem()} respectively at line 17 by the consumer, and at line 34 by another producer. The solution also includes code for accessing the ring buffer using the \texttt{head} and \texttt{tail} variables. The code for the consumer follows the same pattern.

\begin{figure}[!ht]
\includegraphics[width=0.8\columnwidth]{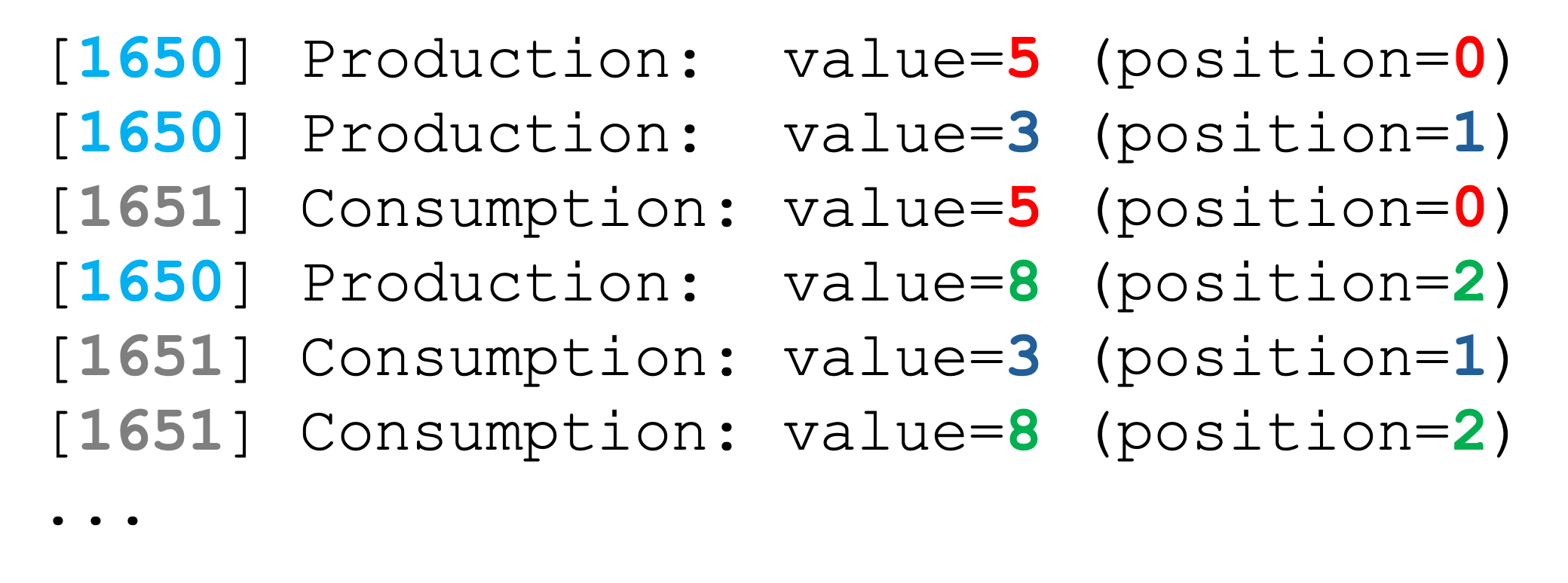}
\vspace{-0.4cm}
\caption{Output of the producer-consumer program.}
\label{fig:example_prod_cons_execution}
\end{figure}

In order to support evaluation checks at the execution stage, the template includes code to print the occurrence of relevant events, such as a production or consumption, and related data, such as the value and position in the ring buffer. Students are invited not to modify these messages, and any change to these lines of code can be detected through the version control system. Figure~\ref{fig:example_prod_cons_execution} shows an excerpt from the output of the solution.

The evaluation checks are performed by a program, as showed in Listing~\ref{alg:procedure_prod_cons}. This procedure iterates over the lines of the output, by checking that productions and consumptions are performed in the right location of the ring buffer (lines 9-11 and 17-19), i.e., at increasing values of the head and the tail of the ring. Moreover, the procedure accumulates the produced and consumed values in two sets (lines 12 and 20). At the end, the procedure checks that the program performed the expected number of productions and consumptions (lines 24-27), and that the consumed and produced values are matching (lines 27-30).

\begin{listing}[!ht]
\caption{Execution checks for the producer-consumer example.}
\label{alg:procedure_prod_cons}
{\footnotesize
\begin{algorithmic}[1]

\Procedure{evaluate\_prod\_cons}{}

\State $produced\_values \gets \emptyset$
\State $consumed\_values \gets \emptyset$
\State $head \gets 0$
\State $tail \gets 0$

\vspace{0.2cm}
    
\ForAll {$line \in output$}

    \vspace{0.2cm}

    \If {$is\_production(line)$}
        \State $pid, value, position = parse(line)$

        \If {$position \ne tail$}
           \State raise error 'Production at wrong location'
        \EndIf

        \State $push(produced\_values, value)$

        \State $tail = (tail + 1) \% SIZE$

    \EndIf
    
    \vspace{0.2cm}   

    \If {$is\_consumption(line)$}
        \State $pid, value, position = parse(line)$

        \If {$position \ne head$}
           \State raise error 'Consumption at wrong location'
        \EndIf

        \State $push(consumed\_values, value)$

        \State $head = (head + 1) \% SIZE$

    \EndIf

    \vspace{0.2cm}   

\EndFor

\vspace{0.2cm}

\If {$|produced\_values|~\ne~TOTAL$ or \\
    \hskip \algorithmicindent  \hskip \algorithmicindent $|consumed\_values|~\ne~TOTAL$}
  \State raise error 'Missing productions/consumptions'
\EndIf

\vspace{0.2cm}

\If {$not ~ equal(produced\_values, consumed\_values)$}
  \State raise error 'Produced/consumed values do not match'
\EndIf

\EndProcedure

\end{algorithmic}
}
\end{listing}

It is important to note that concurrency bugs cannot be caught by just executing the program. For example, if the student omits \texttt{Wait\_Sem()} and \texttt{Signal\_Sem()} on the \texttt{MUTEX\_P} semaphore, concurrent accesses may corrupt the ring buffer (i.e., a race condition). 
In this scenario, the program may still be able to correctly write to the ring buffer by chance. Such a bug would only rarely manifest over a large set of executions, without any guarantee. Many other subtle mistakes, such as the missing allocation or initialization of the semaphores (e.g., \texttt{MUTEX\_P} needs to be initially set to $1$ is order to behave as a mutex), can result in an apparently correct behavior.

\begin{listing}[!t]
\begin{minted}[autogobble,linenos,breaklines,fontsize=\scriptsize]{yaml}
rules:
  - id: semaphore-allocation
    message: "Wrong number of allocated semaphores. This exercise requires 4 semaphores: \"space available\", \"message available\", and two more for mutual esclusion among multiple producers and multiple consumers, respectively."
    patterns:
      - pattern: sem_id = semget($KEY, $NUM_SEM, $FLAGS);
      - metavariable-pattern:
          metavariable: $NUM_SEM
          patterns:
            - pattern-not: "4"

  - id: semaphore-initialization
    message: "Wrong initialization of semaphores. The semaphore \"space available\" must be initialized with the number of buffers (10), the semaphore \"message available\" must be initialized with 0, and the semaphores for mutual exclusion must be initialized with 1."
    patterns:
      - pattern: |
          queue_requests* initialization(...) {
            ...
            semctl($SEMID, $SEM, SETVAL, $INIT);
            ...
          }
      - metavariable-pattern:
          metavariable: $INIT
          patterns:
            - pattern-not: "10"
            - pattern-not: "0"
            - pattern-not: "1"

  - id: producer_synchronization
    message: "The synchronization in the producer is not correct. Use the functions Wait\_Sem() and Signal\_Sem() to apply the producer-consumer algorithm."
    patterns:
      - pattern: |
          void insert_request(...) {
            ...
          }
      - pattern-not: |
          void insert_request(...) {
            ...
            Wait_Sem(...,$SPACE_AVAILABLE);
            Wait_Sem(...,$MUTEXP);
            ...
            Signal_Sem(...,$MUTEXP);
            Signal_Sem(...,$MESSAGE_AVAILABLE);
            ...
          }
\end{minted}
\caption{Static analysis rules for the producer-consumer example.}
\label{alg:prod_cons_static_analysis}
\end{listing}

Static analysis evaluates whether the submitted code follows the expected patterns. Listing~\ref{alg:prod_cons_static_analysis} shows tailored static analysis rules using Semgrep \cite{semgrep}, a popular open-source static analyzer. It provides an easy domain-specific language, based on YAML, to write rules for detecting patterns in source code. This example includes 3 rules, respectively for checking that: (1) the program actually allocates $4$ semaphores with \texttt{semget()}; (2) semaphores were initialized with correct values ($SPACE\_AVAILABLE = 10$, $MESSAGE\_AVAILABLE = 0$, $MUTEX\_P = MUTEX\_C = 1$); and, (3) the functions \texttt{Wait\_Sem()} and \texttt{Signal\_Sem()} were invoked in the right order.

The rules contain a \emph{pattern} attribute, which includes a piece of C code that the tool will look for in the source code. Our rules look for known places in the code that need to be completed by the students, such as the \emph{insert\_request} and the \emph{initialization} functions in the template code. When a match is found, the rules checks additional patterns listed in the \emph{pattern-not} attribute. In our case, the \emph{pattern-not} attribute lists the expected correct value for the input parameters of \texttt{semget} and \texttt{semctl}, and the expected code for \texttt{Wait\_Sem()} and \texttt{Signal\_Sem()}. If none of the \emph{pattern-not}s is matched, the tool raises an error, showing the information in the \emph{message} attribute of the rule.

{
\footnotesize
\begin{table*}[ht]
\caption{Systems programming assignments over the semester.}
\label{tab:exercises}
\vspace{-0.3cm}
\begin{tabularx}{\textwidth}{@{}>{\hsize=.35\hsize}X>{\hsize=0.9\hsize}X>{\hsize=1.75\hsize}X@{}}
\toprule
\textbf{Assignment} & \textbf{Exercise} & \textbf{Description} \\
\midrule
\textit{A0. Introduction}                                 & Hello world                                        & Print a simple "hello   world" message. Introduction to the Linux shell, Make, and Git. \\
\midrule
\multirow{7.5}{*}{\textit{A1. Semaphores}}                  & Parallel   Computing over a Large Vector  (mutex)     & Parallel search of minimum value in a vector, saving the value in a shared buffer in mutual exclusion. \\ \cmidrule{2-3}
                                                      & Parallel   Computing over a Large Vector  (prod/cons)     & As the previous exercise, but using the single-buffer producer-consumer scheme to save the value. \\ \cmidrule{2-3}

                                                      & Producer-Consumer on a Pair of Buffers              & Producer-Consumer   using a variable to track the state of each buffer: empty, full, or "in   use". Buffers "in use" are simulated by slowing down   operations with sleep(). \\ \cmidrule{2-3}
                                                      & Reader-Writer on a Data Structure                  & Reader-Writer on a buffer containing a complex data type (instead of a simple integer). \\ \cmidrule{2-3}
                                                      & I/O   Scheduler Simulation using a Circular Buffer & Producer-Consumer using a   circular queue of buffers. \\
\midrule
\multirow{5}{2cm}{\textit{A2. Monitor Construct}}           & I/O Scheduler Simulation using a   Circular Buffer & Producer-Consumer using a   circular queue of buffers. \\ \cmidrule{2-3}
                                                      & Producer-Consumer with Multiple Consumptions       & Producer-Consumer,   the Consumer consumes two values for each atomic operation. \\ \cmidrule{2-3}
                                                      & Reader-Writer on a Data Structure                  & Reader-Writer on   a buffer containing a complex data type, using the Monitor Construct. \\ \cmidrule{2-3}
                                                      & Producer-Consumer   with Priority                  & Producer-Consumer, with two   priority levels for suspending and waking-up Producers. \\
\midrule
\multirow{5}{*}{\textit{A3. Threads}}                     & Thread-safe   Stack Data Structure                 & Producer-Consumer   on a Stack data structure, with methods push(), pop(), and size(). The size()   method has to be in mutual exclusion despite it only reads data. \\ \cmidrule{2-3}
                                                      & Reader-Writer on Multiple Monitor Objects          & Reader-Writer on   a vector of Monitor objects, with each object is synchronized separately. \\ \cmidrule{2-3}
                                                      & Producer-Consumer   with Priority                  & Producer-Consumer, with two   priorities for waking-up suspended Producers.  \\
\midrule
\multirow{9}{2cm}{\textit{A4. Message Queues}}              & Load Balancing                                     & Client-Server, with an   intermediate Balancer process that forwards each message to one of three   Servers in round-robin.    \\ \cmidrule{2-3}
                                                      & Multiple Synchronous Servers                       & Client-Server,   with multiple servers each with a dedicated message queue, and two additional   shared message queues for handshake synchronization.   \\ \cmidrule{2-3}
                                                      & Dependency Graph                                   & Parallel   computation of an arithmetic formula, with message passing to share operands   and results. \\ \cmidrule{2-3}
                                                      & Distributed   Registry                             & Client-Server, with multiple   servers each with a dedicated message queue, with a Registry process to track   and retrieve the queue identifiers.  \\
\midrule
\multirow{6}{1.6cm}{\textit{A5. Threads with Message Queues}} & Multithread Server                                 & Client-Server, using dedicated   threads to handle each request.   \\ \cmidrule{2-3}
                                                      & Remote Procedure Call                              & Client-Server   with RPC pattern, using proxy methods to make requests through a message   queues, and worker threads to execute the calls on the Server.   \\ \cmidrule{2-3}
                                                      & Aggregator                                         & Client-Server with an   intermediate ``aggregator'' process. A thread in the aggregator receives a message and shares it with other threads (Reader-Writer), which in turn send the value to server processes. \\
\bottomrule
\end{tabularx}
\end{table*}
}

\subsection{General criteria for evaluation checks}
\label{subsec:general_criteria}

Static and dynamic evaluation checks have been systematically applied on all exercises of the class. The following general criteria were adopted to decide which checks to perform, and in which stage of the evaluation (execution or static analysis).

The checks at execution are responsible for evaluating:

\begin{itemize}[label={}, leftmargin=*]

\item \textbf{Progress}. The program should be able to complete the expected operations. The template code provides business logic to iterate the operations of the program, such as, productions and consumptions, client-server interactions, and remote procedure calls. The template code includes \texttt{printf()}s, in order to check with dynamic analysis the number of iterations that were actually performed.

\item \textbf{Values}. The output of the program includes values processed by the program (typically, small random integer numbers). Dynamic analysis checks that these values are correctly exchanged and processed across multiple processes.

\item \textbf{Order}. Dynamic analysis checks that the program follows order relations between events, according to synchronization patterns. For example, a value can be consumed only after it is produced, or it can be read only after it is written.

\end{itemize}

Static analysis checks focus on evaluating:

\begin{itemize}[label={}, leftmargin=*]

\item \textbf{Initialization}. Many synchronization algorithms require careful data initialization, including semaphores, counters, array indexes, and state variables.

\item \textbf{Synchronization algorithms and conditions}. Exercises are designed to cover several types of synchronization algorithms, with different flavors of producer-consumer and reader-writer, such as: with single or multiple buffers, using head/tail pointers or an array of state variables for each buffer; with semaphores or the monitor construct; with multiple processes or threads. The exercises require the use of the correct APIs at the right place, such as \texttt{\{Wait,Signal\}\_Sem()} for semaphores, and  \texttt{pthread\_\{lock,} \texttt{unlock,} \texttt{cond\_}\texttt{wait,} \texttt{cond\_} \texttt{signal\}()} for the monitor construct with threads. Moreover, the checks evaluate the use of variables at synchronization conditions, such as, head/tail to suspend a producer on a condition variable. In the case of critical sections, the checks evaluate that the positioning of primitive calls does not make the section too large, which would unnecessarily slow down the program, or too small, which would expose data to race conditions.

\item \textbf{Resource management}. OS resources for concurrent programming include UNIX semaphores, shared memory, and message queues \cite{kerrisk2010linux,kay2003unix}. The scope of these resources is not limited to the lifecycle of an individual process. Thus, they need to be carefully allocated and initialized by only one process, and de-allocated only once all processes have completed. This involves the correct use of \emph{resource keys} to: create a new resource or to retrieve an existing one; use distinct keys when the program requires multiple instances of the same resource; use key-less resources only when appropriate (e.g., resources to be shared only between a process and its children ones). Static analysis assesses the correct use of systems calls for resource management.

\item \textbf{Messaging}. Exercises include several cases of message passing using queues, such as: blocking receivers and non-blocking senders; both blocking senders and receivers (\emph{handshake}); indirect addressing using the same mailbox for multiple receivers; direct addressing using typed messages; protocol state machines. Static analysis assesses the correct use of systems calls for messaging.

\end{itemize}

A limitation of these static analysis checks is that they are highly accurate for assignments that are sufficiently constrained. For example, in the producer-consumer scheme, there must be semaphores initialized to 0 and to the number of buffers, respectively. Instead, assignments that allow different design choices (e.g., different synchronization schemes) may not be checked with high accuracy. In this approach, we consider assignments where the synchronization scheme and the structure of the program are constrained. Still, static analysis checks are flexible enough to allow variants of the correct solution (e.g., with different order of non-synchronization statements).

\section{Case study}
\label{sec:case_study}

\begin{figure*}[!htbp]
\includegraphics[width=\textwidth]{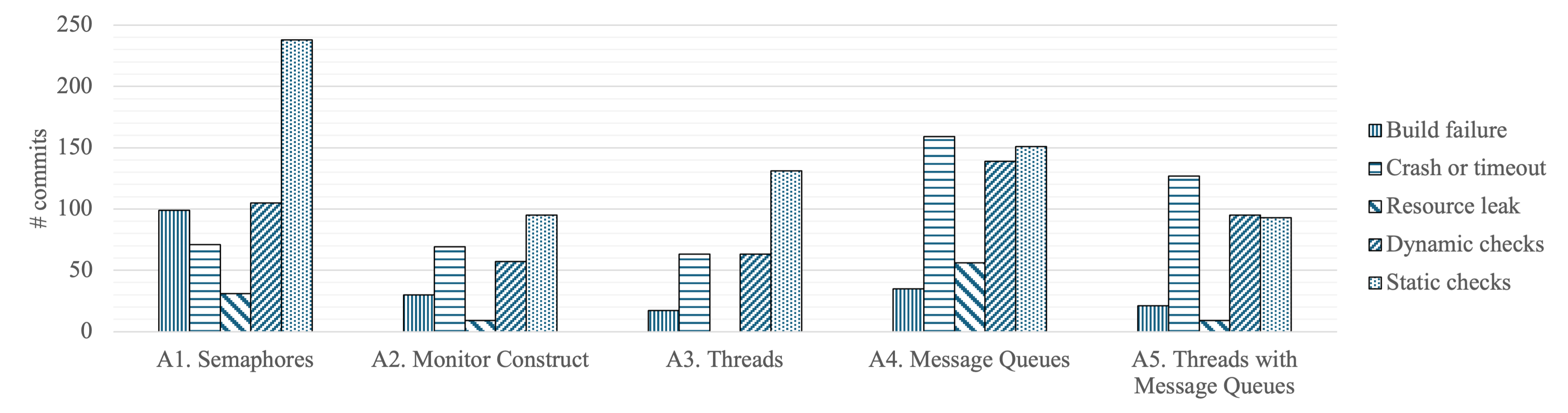}
\caption{Detection of commits with errors, by type of evaluation check.}
\label{fig:checks}
\end{figure*}

We present an experience report from an Operating Systems (OS) class at the \theinstitution. The class is attended by third-year undergraduate students of the BSc degree in Computer Engineering, with basic knowledge of C and C++ programming from previous classes. The class introduces students to basic OS concepts, including OS abstractions, CPU scheduling, memory and I/O management, filesystems, and virtualization.

The class includes both theory lectures and systems programming lectures, in order to apply theory in a practical context. The theory lectures are scheduled at the first and last months. In the middle, the programming lectures present concurrency schemes based on shared memory and message passing, and how to implement them using UNIX System V IPC \cite{kerrisk2010linux,kay2003unix}. Every week, a set of exercises is assigned to students for practice. The students can solve the exercises both during a 2-hours session in the classroom, and in their own time. The final exam consists of a programming exercise of the same type as the exercises of these assignments. 

The systems programming exercises of the case study are listed in Table~\ref{tab:exercises}. 
Each assignment consists of a set of programs. In total, there are 5 assignments, with an additional introductory assignment at the beginning about basic tools. The exercises address typical patterns of concurrent systems in simplified form, such as, multi-threaded servers and remote procedure calls. The template code allows students to focus on synchronization, messaging, and OS resource management. In some cases, the same topic is assigned across multiple exercises in different forms, in order to emphasize differences between affine concepts, such as, the same producer-consumer implemented with semaphores and then with the monitor construct, or with processes and then with threads. 

The template code for the exercises is publicly available \cite{es-intro,es-semafori,es-monitor,es-threads,es-code,es-server}. The template code includes a hidden \texttt{.test} folder, with scripts and configuration files for the static and dynamic checks, which are executed using GitHub actions \cite{decan2022use}. Assignments are managed using GitHub Classroom \cite{hecht2023github}, where feedback is posted in a conversation thread of a pull request \cite{poc-autograder}. Static checks were implemented with Semgrep, and dynamic checks with Bash and Perl scripts.

On average, $120$ students joined each assignment, of which $95\%$  submitted at least one commit, and of which $54\%$ completed all exercises of the assignment, by passing all the evaluation checks. They were allowed to make multiple resubmissions. Only in a few cases, students reported false positives from the evaluation checks, which were related to synchronization APIs missed by our static analysis rules (e.g., \texttt{pthread\_cond\_broadcast()}) and unusual control flow structures for synchronization conditions with \texttt{if}/\texttt{else}. These issues were promptly fixed to allow correct programs to pass.

Figure~\ref{fig:checks} shows the distribution of submitted commits with an error, as detected by our evaluation process. The distribution is divided by type of evaluation check, including build failures, crash or timeout of the program, leak of OS resources, dynamic checks on the outputs, and static checks.

We observe that in the first assignment, the number of build failures is relatively high, and becomes lower as the students become familiar with the automatic evaluation process, getting the habit to build the code locally before submitting it. 

The \emph{generic} checks (i.e., checks that are the same across assignments), which include checks for crashes, timeouts, and OS resource leaks, are between $19\%$ and $40\%$. They detected fewer errors than the \emph{specific} checks (i.e., checks that were written specifically for an assignment), which include dynamic and static checks, and which are between $54\%$ and $71\%$. This result remarks the importance of using specific checks to analyze in depth the behavior of the program and its internals. 

In all five assignments, a significant number of faulty commits (between $27\%$ and $44\%$) were detected by static analysis. This percentage was higher for the first assignment, since it is the first time that students practice systems programming and system calls for UNIX System V IPC. 
The number of faulty commits detected by dynamic checks (between $19\%$ and $28\%$) is similar or lower than those detected by static checks. This result shows that dynamic checks (which apply the same approach of traditional automatic evaluation systems) are not sufficient in the case of systems programming. Since concurrency and resource management bugs are difficult to reproduce, they tend to escape checks performed by running the program, and they can only be detected by a deeper analysis of the internals of the program. In our experience, static analysis rules were effective at identifying a significant amount of systems programming errors.

\section{Conclusion}
\label{sec:conclusion}

In this paper, we presented an experience in automatic evaluation of systems programming exercises. Systems programming poses special challenges because of synchronization and resource management bugs. Traditional automatic evaluation systems struggle with these bugs, as they can only rely on the execution of test cases. We presented the design of systems programming exercises, and how to tailor static analysis rules to complement the evaluation process. Static analysis was able to identify a large number of faulty submissions, thus providing valuable feedback to students.

\begin{acks}
This work has been partially supported by MUR PRIN 2022, project FLEGREA, CUP E53D23007950001 (\url{https://flegrea.github.io}). I am grateful to my colleagues at the Federico II University for the collaboration on teaching Operating Systems classes.
\end{acks}

\onecolumn
\begin{multicols}{2}
\bibliographystyle{ACM-Reference-Format}
\bibliography{bibliography}
\end{multicols}

\end{document}